\documentstyle[12pt]{article}

\begin{document}
\title{SUSY model with R-parity violation, longlived charged slepton and 
quasistable supermatter}
\author{N.V.Krasnikov \\Institute for Nuclear Research, Moscow 117312\\}

\date{January, 1996}
\maketitle
\begin{abstract}
We construct two SUSY $SU(3) \otimes SU(2)_{L} \otimes U(1)$ electroweak models 
with superweak R-parity violation. The scale of R-parity violation in one 
of the models is determined by the Majorana mass of neutrino and it is 
very small that leads to the existence of longlived ($T \geq O(10^{-4})sec $) 
lightest superparticle. If lightest superparticle is righthanded charged 
slepton that can be realized within gaugino dominated scenario then the 
phenomenology of such model differs in a drastic way from the standard SUSY 
phenomenology, in particular, longlived charged sleptons can form bound 
states with ordinary matter - quasistable supermatter (SUSY analogs of 
mu-atoms and muonium). We discuss possible manifestations of the existence 
of such longlived charged particle at LEP2, TEVATRON and LHC. We also 
construct $SU(3) \otimes SU(2)_{L} \otimes SU(2)_{R} \otimes U(1)$ 
model with Majorana mass and superweak $R$-parity violation 
\end{abstract}
\begin{flushright}
Preprint INR96/0916
\end{flushright}

\newpage

Supersymmetric electroweak models offer the simplest solution of the 
gauge hierarchy problem \cite{1}-\cite{4}. In real life supersymmetry has 
to be broken and the masses of superparticles have to be lighter than 
$O(1)$Tev provided the supersymmetry solves the gauge hierarchy problem 
 \cite{4}. Supergravity gives natural explanation of the 
supersymmetry breaking, namely, an account of the supergravity breaking in 
hidden sector leads to soft supersymmetry breaking in observable sector 
\cite{4}. For the supersymmetric extension of the Weinberg-Salam model 
  soft supersymmetry breaking terms usually consist of the gaugino mass 
terms, squark and slepton mass terms with the same mass at Planck scale and 
trilinear soft scalar terms proportional to the superpotential \cite{4}. 
Other standard assumption is that the "Minimal Supersymmetric Standard Model" 
(MSSM) conserves R-parity  \cite{4}. In the MSSM the superpartners 
of the standard model states are $R$-odd while all standard model states 
are $R$-even. As a consequence of $R$-parity conservation the new 
supersymmetric states can only be produced in pairs and supersymmetric 
states can't decay into ordinary particles so the lightest superparticle 
(LSP) is stable. This has very important effect on the supersymmetry search. 
Namely, experimental searches for new supersymmetric particles are based on 
pair production of superparticles and the typical SUSY signature involves 
missing $p_{T}$ momentum as a signal for SUSY particles production.
However $R$ need not to be conserved in supersymmetric extensions of the 
Weinberg-Salam model. Moreover the most general form for the renormalized 
superpotential in MSSM contains terms which explicitly violate $R$ that 
leads in general to the nonconservation of lepton number and proton decay.
The phenomenology of the models with $R$-violation has been discussed 
in refs.\cite{5}-\cite{13} in the assumption that the lightest 
superparticle is electrically neutral ($U(1)$ gaugino or neutral Higgsino). 
For longlived gaugino missing $p_{T}$ momentum signature dissapears, 
however typically such models predict the excess of multileptons due to 
superparticle decays that allows to discover SUSY even for the case of 
$R$-parity breaking. For the lightest superpaticle  decaying outside 
the detector the phenomenology coincides with the standard one. 

In this paper we construct two SUSY $SU(3) \otimes SU(2)_{L} \otimes U(1)$ 
models with $R$-parity violation. In the first model the $R$-parity 
violation is due to the exchange of the heavy singlet superfield.
In the second model the scale of $R$-breaking is determined by the neutrino 
Majorana mass and as a consequence the lightest superparticle is longlived
$T_{lsp} \geq O(10^{-4})$ sek. We assume that the lightest superparticle is 
charged righthanded slepton. The situation when the lightest superparticle 
is righthanded slepton is realized within supergravity motivated models 
for the case of gaugino dominated scenario (at GUT scale slepton and squark 
masses are small compared to gaugino masses). Longlived charged slepton 
can form bound states with ordinary leptons and nuclei - quasistable 
supermatter (analogs of muonium and mu-atoms). We discuss the 
manifestations of the existence of quasistable charged slepton at LEP2, 
TEVATRON and LHC. For the case when the lightest superparticle is 
neutralino the phenomenology coincides with the standard one that is not 
very interesting. We also construct 
$SU(3) \otimes SU(2)_{L} \otimes SU(2)_{R} \otimes U(1)$ model 
with naturally small $R$-violation. 

Let us add to the minimal supersymmetric $SU(3) \otimes SU(2)_{L} \otimes U(1)$  
model two singlet superfields $\Phi_{1}$ and $\Phi_{2}$. The superpotential 
of the model has the form (here we restricted ourselves to the third 
generation)
\begin{equation}
W = W_{m} + W_{\Phi} + W_{\mu} \ ,
\end{equation}
\begin{equation}
W_{m} = h_{t}QH_{1}\bar{t} + h_{b}QH_{2}\bar{b} + 
h_{\tau}LH_{2}\bar{\tau} \ ,
\end{equation}    
\begin{equation}
W_{\Phi} = \lambda_{1}\Phi_{1}H_{1}H_{2} + \lambda_{2} \Phi_{2}LH_{1} + 
M \Phi_{1}\Phi_{2} \ ,
\end{equation}
\begin{equation}
W_{\mu} = \mu H_{1}H_{2} 
\end{equation} 
Here $Q = (t,b)_{L}$, $L = (\nu, \tau)_{L}$, $H_{1} = (H_{11}, H_{12})$ , 
$H_{2} = (H_{21}, H_{22})$, $\bar{t} = t_{R}^{c}$, $\bar{b} = b_{R}^{c}$ , 
$\bar{\tau} = {\tau}^{c}_{R}$, $H_{1}H_{2} = {\epsilon}^{ij}H_{1i}H_{2j}$. 
The superpotentials $W_{m}$ and $W_{\Phi}$ are invariant under 
the transformations
\begin{equation}
H_{1} \rightarrow exp(i\alpha_{1})H_{1} \ ,
\end{equation}
\begin{equation}
H_{2} \rightarrow exp(i\alpha_{2})H_{2} \ ,
\end{equation}
\begin{equation}
L \rightarrow exp(-i\alpha_{1}-2i\alpha_{2})L \ ,
\end{equation}
\begin{equation}
Q  \rightarrow \exp(i\alpha_{4})Q \ ,
\end{equation}
\begin{equation}
\bar{t} \rightarrow exp(-i\alpha_{4}-i\alpha_{1}) \bar{t} \ ,
\end{equation}
\begin{equation}
\bar{b} \rightarrow exp(-i\alpha_{2} - i\alpha_{4}) \bar{b}  \ ,
\end{equation}
\begin{equation}
\bar{\tau} \rightarrow exp(2i\alpha_{1}) \bar{\tau} \ ,
\end{equation}
while the term $W_{\mu}$ breaks the symmetry (5-6) in a soft way. 
The symmetry (5-11) prohibits renormalizable $R$-parity violating terms
\begin{equation}
\delta W = \lambda_{ijk}L_{i}L_{j}\bar{l}_{k}  +  
\lambda_{ijk}^{'}L_{i}Q_{j}\bar{d}_{k} + 
\lambda_{ijk}^{''}\bar{u}_{i}\bar{d}_{j}\bar{d}_{k} 
\end{equation}
Here $L_i$, $\bar{l}_{j}$,  $Q_j$, $\bar{u}_{j}$, $\bar{d}_{k}$ are lepton 
doublets, charged lepton singlets, quark doublets, up quark singlets, 
down quark singlets respectively. After integration over the heavy 
superfields $\Phi_{1}$ and $\Phi_{2}$ we find an effective superpotential 
violating $R$-parity  
\begin{equation}
W_{R} = k H_{1}H_{2}LH_{1} \ ,
\end{equation}
where
\begin{equation}
k = \frac{\lambda_{1}\lambda_{2}}{M}
\end{equation}
The smallness of R-parity violation in considered model is due to assumed 
big mass $M$ of $\Phi_{1}$ and $\Phi_{2}$ singlets.

Consider now electroweak $SU(3) \otimes SU(2)_{L} \otimes U(1)$ supersymmetric 
model with righthanded neutrino. The superpotential of the model for the 
third generation has the form 
\begin{equation}
W = W_{m} + W_{1} \ ,
\end{equation} 
\begin{equation}
W_{m} = h_{t}QH_{1}\bar{t} + h_{b}QH_{2}\bar{b} + 
h_{\nu}LH_{1}\bar{\nu} + h_{\tau}LH_{2}\bar{\tau} \ , 
\end{equation}
\begin{equation}
W_{1} = \mu H_{1}H_{2} + \frac{M_{\nu} \bar{\nu}\bar{\nu}}{2} + 
\lambda H_{1}H_{2}\bar{\nu}
\end{equation}
Here $\bar{\nu} = \nu^{c}_{R}$.   
The superpotential $W_{m}$ conserves $R$-parity while the term 
$\lambda \bar{\nu} H_{1}H_{2}$ in the superpotential 
$W_{1}$ violates $R$-parity. After integration over neutrino 
superfield $\bar{\nu}$ we find an effective $R$-parity violating 
superpotential (13) with coefficient
\begin{equation}
k = \frac{\lambda h_{\nu}}{M_{\nu}}
\end{equation} 
The smallness of $R$-parity violation is due to big neutrino Majorana mass 
$M_{\nu}$. We shall consider the most interesting case when the righthanded 
$\tau$-slepton is the lightest superparticle. Such situation takes place 
in the so called gaugino dominated scenario \cite{14}-\cite{16}. Really, 
in supergravity motivated models standard assumption is that at GUT scale 
soft supersymmetry breaking terms have very simple structure: all squark and 
slepton masses coincide, gaugino masses also coincide and the trilinear 
soft supersymmetry breaking term is proportional to the superpotential. 
In such scenario the lightest superpartner of quarks and sleptons is the 
righthanded slepton \cite{17}-\cite{18}. For the case when Yukawa coupling 
constants of leptons are negligible all righthanded sleptons are degenerate 
in mass. In terms of soft supersymmetry breaking parameters $m_{0}$ 
(common squark and slepton mass at GUT scale) and $m_{1/2}$ 
(common gaugino mass) righthanded slepton mass is given by the 
formula \cite{18}
\begin{equation}
\tilde{m}^{2}_{E_{R}} = m^2_0 + 0.14m^2_{1/2} -0.22\cos(2\beta)M^2_Z \,
\end{equation}    
For the case of the third generation the $\tau$-lepton Yukawa coupling is 
the biggest one among leptons and an account of nonzero Yukawa coupling 
effects leads to the decrease of the corresponding righthanded slepton 
masses \cite{18}, so the righthanded $\tau$-slepton is the lightest 
sparticle among squarks and sleptons. For gaugino masses an account of 
evolution from GUT scale to observable electroweak scale leads to the 
formula \cite{18}
\begin{equation}
M_{i} = \frac{\bar{\alpha}_{i}(M_Z)}{\alpha_{GUT}}m_{1/2}
\end{equation}
Gaugino associated with $U(1)$ gauge group is the lightest sparticle 
among gauginos and numerically its mass is given by the formula
\begin{equation}
M_{1} \approx 0.43m_{1/2}
\end{equation}
Comparing formulae (15) and (21) we find that for gaugino dominated scenario 
\cite{14}-\cite{16} ( $m_{0} \ll m_{1/2}$, to be precise for 
$m_{0} \leq 0.17m_{1/2}$) the righthanded $\tau$-slepton is the 
lightest superparticle. In principle the higgsino can be the lightest 
superparticle. However if we require that radiative corrections to the 
tree level effective potential give the correct electroweak vacuum then 
it is possible to determine numerically the higgsino mass in terms of 
$m_{0}$ and $m_{1/2}$. We have checked that for the gaugino dominated 
scenario the righthanded slepton is really the lightest superparticle. 
It should be noted that gaugino dominated scenario allows to solve 
\cite{19} the problem with flavor changing neutral 
currents arising due to nonuniversal squark mass matrix.  
For the case $m_{0} \geq 0.17m_{1/2}$ we reproduce standard and not 
very interesting for us case when $U(1)$ gaugino  (its mixture 
with other neutralino) is the lightest superparticle. We shall assume 
in this paper that the lightest superparticle is the charged righthanded 
$ \tau $-slepton. For the model with exact $ R$-parity the existence 
of stable electrically charged particle contradicts to the experimental 
data on abundances of anomalous  superheavy isotopes \cite{20}-\cite{22}, 
however for models with explicitly broken  $R$-parity the existence of 
longlived charged superparticle does not contradict to arguments based on 
abundances of heavy isotopes. 
  
In considered models righthanded $\tau$-slepton $\bar{\tau}$ will decay 
into lefthanded $\tau$-lepton and $\tau$-neutrino.        
After electroweak symmetry breaking ( $<H_{1}> \ne 0$ , $<H_2> \ne 0$) we 
find that the effective Lagrangian describing the righthanded 
$\tau$-slepton decay into neutrino and $\tau$-lepton has the form
\begin{equation}
L_{\bar{\tau} \rightarrow \tau {\nu}_{\tau}} = h\tilde{\tau}^{+}_{R}
\tau_{L}\nu_{L} + h.c. \,,
\end{equation}
where
\begin{equation}
h = \frac{k h_{\tau}<H_1><H_2>}{\mu}
\end{equation}
In the second model the smallness of the neutrino mass is due to 
see-saw mechanism \cite{23}, namely
\begin{equation}
m_{\nu_{\tau}} = \frac{(<H_1>h_{\nu})^2}{M_{\nu}}
\end{equation}
So the coupling constant $h$ can be rewritten in the form
\begin{equation}
h = \frac{\lambda m_{\tau}m_{\nu}}{\mu m_{D\nu}} \,,
\end{equation}
where $m_{\tau} = h_{\tau}<H_2>$ and $m_{D\nu} = h_{\nu}<H_1>$.
For the Lagrangian (22) the decay width of the righthanded $\tau$-slepton 
into $\tau$-lepton and $\nu_{\tau}$-neutrino is determined by the formula 
\begin{equation}
\Gamma(\tilde{\tau}_{R} \rightarrow \tau \nu_{\tau}) =
\frac{h^{2}M_{\tilde{\tau}_{R}}}{16\pi}
\end{equation}
From upper cosmological bound $m_{\nu} \leq 10 $ ev on neutrino mass 
in the assumption that 
$m_{D\nu} \sim m_{\tau}$ and $\mu \sim 100$ Gev we find that the lifetime 
of the righthanded $\tau$-slepton is 
\begin{equation}
T_{\tilde{\tau}_R} \sim O(10^{-4}(\lambda)^{-2}) \, sec
\end{equation}
For the first model in the suggestion that $\lambda_{1} \sim 1$, 
$\lambda_{2} \sim 1$ and the $\Phi_{1}$ and $\Phi_{2}$ supersinglet 
mass is determined by GUT scale $M \sim M_{GUT} \sim 10^{16}$ Gev. 
we find that righthanded slepton lifetime is $T_{lsp} \sim $ several years.

Therefore we  find that in constructed models the lightest superparticle 
is longlived charged particle (it is not decay within detector) that 
leads to the different strategy for the search for supersymmetry at 
supercolliders. Such longlived charged particles can form boundary 
states with nuclei and electrons - quasistable supermatter 
(analogs of mu-atoms and muonium). If neutralino is the lightest 
superparticle then it is also longlived and decays outside the detector 
so the supercollider phenomenology coincides with the standard one that 
is not very interesting. 
 
Consider briefly the phenomenology of such longlived particles. At LEP2 
the cross section of the reaction
\begin{equation} 
e^{+}e^{-} \rightarrow {\tilde{\tau}_{R}}^{+}{\tilde{\tau}_{R}}^{-}
\end{equation}
is well known \cite{24}-\cite{25} and it is equal to 0.064 pb(0.033 pb) for 
$M_{\tilde{\tau}} = 85 Gev(90Gev)$. For the Luminosity $L = 500 pb^{-1}$ 
we expect 32 events for $M_{\tilde{\tau}} =85 Gev$ and 16 events for 
$M_{\tilde{\tau}} = 90 Gev$. It is not difficult to distinguish 
the reaction (28) from the standard reaction $e^{+}e^{-} \rightarrow 
{\mu}^{+}{\mu}^{-}$. By the measurement of the momentum $\vec{p}$ of the 
charged particle in magnetic field it is possible to determine its mass 
using the standard formula $M = \sqrt{\frac{s}{4} - {\vec{p}}^2}$.
So we conclude that LEP2 will be able to discover longlived 
righthanded $\tau$-sleptons with the masses up to 90 Gev. At TEVATRON 
and LHC due to the smallness of $R$-parity violating interaction 
(13) SUSY particles will produce in pairs like in standard scenario.  
In the final states (after decays of squarks, qluino, wino, 
sleptons and neutralino into ordinary particles and righthanded 
$\tau$-slepton) we shall have two longlived charged particles. Again 
it is possible to distinguish such particles from muons by the measurement 
of their momentum, as a consequence of nonzero mass of righthanded 
slepton we shall have missing $E_{T}$ if we misidentify righthanded 
slepton as muon.  
\footnote{ Detailed discussion of the possibility to detect charged 
longlived particles at LEP2, TEVATRON and LHC will be given elsewhere}    
 In our second model (15) with explicit $R$-violation we put "by hands" 
possible renormalizable $R$-violating terms (12). In general it is not 
clear why such terms are absent and $R$-violation is small. 

To overcome this shortcoming consider   
$SU(3) \otimes SU(2)_{L} \otimes SU(2)_{R} \otimes U(1)$ 
generalization of our $SU(3) \otimes SU(2)_{L} \otimes U(1)$ model (15) with 
righthanded neutrino. Superquark multiplets and superlepton multiplets 
have the following transformation rules under 
$SU(2)_{L} \otimes SU(2)_{R} \otimes U(1)$ gauge group:
\begin{equation}
Q_{L} \equiv (2,1,1/3); \, \bar{Q_{R}} \equiv (1,2,-1/3) \,,
\end{equation}
\begin{equation}
\Psi_{L} \equiv (2,1,-1); \, \bar{\Psi_{R}} \equiv (1,2,1)
\end{equation}
The superhiggs sector of the model consists of a superhiggs bidouplet   
$\Phi(2,2,0)$ which corresponds to two superhiggs doublets $H_1$ and $H_2$ of 
the standard $SU(3) \otimes SU(2)_{L} \otimes U(1) $ model. In the minimal 
left-right model \cite{26}-\cite{27} superhiggs sector consists also of 
superhiggs triplets 
$\Delta_{L}(3,1,2) \oplus \bar{\Delta}_{L}(3,1,-2)$ and 
$\Delta_{R}(1,3,2) \oplus \bar{\Delta}_{R}(1,3,-2)$. Nonzero vacuum 
expectation values of supermultiplets  
$\Delta_{R}(1,3,2) \oplus \bar{\Delta}_{R}(1,3,-2)$ 
lead to $ SU(2)_{L} \otimes SU(2)_{R} \otimes U(1) \rightarrow
SU(2)_{L} \otimes U(1)$ breaking, whereas nonzero vacuum expectation values 
of superhiggs bidoublet $\Phi(2,2,0)$ are responsible for $SU(2)_{L} \otimes 
U(1)$ breaking and nonzero fermion masses. However for such 
superhiggs structure it is impossible to write down gauge invariant  
$R$-violating terms in the superpotential containing an odd number of 
supermatter superfields, so the $R$-parity for such superhiggs structure 
is conserved automatically, that is not interesting for us. 
To break $SU(2)_{R}$ gauge group we shall use superhiggs doublets 
$H_{R} \equiv (1,2,-1) \oplus \bar{H}_{R} \equiv (1,2,1)$  and
$H_{L} \equiv (2,1,-1) \oplus \bar{H}_{L} \equiv (2,1,1)$. 
Nonzero vacuum expectation value of the superhiggs $H_{R} \oplus 
\bar{H}_{R}$ leads to 
$SU(2)_{L} \otimes SU(2)_{R} \otimes U(1) \rightarrow SU(2)_{L} \otimes U(1)$
breaking. The terms in the superpotential responsible for nonzero 
neutrino Majorana mass and $R$-violation read 
\begin{equation}
W_{1} = \lambda_{1}\bar{\Psi}_{R}H_{R}\varphi + \lambda_{2}\Phi\Phi\varphi 
+  \frac{M_{1}{\varphi}^2}{2}
\end{equation}
Here $\varphi$ is $SU(2)_{L} \otimes SU(2)_{R} \otimes U(1)$ singlet 
superfield. The superpotential (31) and supermatter terms $M\varphi M$ 
$(M \equiv (Q_{L}, \bar{Q}_{R}, \Psi_{L}, \bar{\Psi}_{R}))$ are 
invariant under the discrete transformations
\begin{equation}
\Phi \rightarrow \exp{(i\frac{\pi}{2})} \Phi \,,
\end{equation}
\begin{equation} 
\varphi \rightarrow  -{\varphi} \,,
\end{equation}
\begin{equation}
(H_{R}, \bar{H}_{R}) \rightarrow 
\exp{(-i\frac{3\pi}{4})}(H_{R}, \bar{H}_{R})\ ,
\end{equation}  
\begin{equation}
M \rightarrow \exp{(-i\frac{\pi}{4})}M \,,
\end{equation}
After integration over heavy superfield $\varphi$ we find the effective 
superpotential
\begin{equation}
W_{1}^{'} = \frac{(\lambda_{1}\bar{\Psi}_{R}H_{R} + \lambda_{2}\Phi\Phi)^2}
{M_{1}}
\end{equation}
describing $R$-parity violation. 
After $SU(2)_{R}$ gauge symmetry breaking neutrino acquires nonzero 
Majorana mass $M_{\nu} = \frac{2{\lambda}^{2}_{1}{<H_{R}>}^{2}}{M_{1}}$. The 
term $\bar{\nu}H_{1}H_{2}$ in the superpotential (17) originates from the 
superpotential (36) with the coefficient $\lambda = \frac{2\lambda_{1}
\lambda_{2}<H_{R}>}{M_{1}}$.
So we see that $SU(2)_{L} \otimes SU(2)_{R} \otimes U(1)$ generalization 
of the standard $SU(2)_{L} \otimes U(1)$ electroweak gauge group allows 
to obtain the smallness of the $R$-violation in a natural way as a 
consequence of big neutrino Majorana mass.

To conclude, we proposed two $SU(3) \otimes SU(2)_{L} \otimes U(1)$ 
 models with small $R$-violation. In gaugino dominated scenario the 
lightest superparticle is the charged righthanded $\tau$-slepton and 
its lifetime is $\geq O(10^{-4})$ sec (for the model with 
righthanded neutrino). It means that such longlived charged particle 
can form bound states with ordinary nuclei and electrons - quasistable 
supermatter (analog of mu-atoms and muonium). 
The supercollider phenomenology differs from the standard one in a drastic 
way, namely, in such model in final states there are two righthanded 
$\tau$-sleptons coming from squark, gluino, chargino, neutralino and slepton 
decays. $SU(2)_{L} \otimes SU(2)_{R} \otimes U(1)$ generalization of 
the standard $SU(2)_{L} \otimes U(1)$ electroweak gauge group allows to 
obtain naturally very small $R$-violation as a consequence of big neutrino 
Majorana mass.   
   
I thank CERN TH Department for the hospitality during my stay at CERN where 
this paper has been started. I am indebted to the collaborators of the INR 
theoretical department and especially to V.A.Matveev and V.A.Rubakov 
for discussions and critical comments. The research described in this 
publication was made possible by Grant 94-02-04474-a of the Russian 
Scientific Foundation.

\end{document}